%% file: ICCAS-manuscript.tex
\documentclass[10pt,twocolumn]{ICCAS}

\usepackage{diagbox}
\usepackage{cite}

\def\cmt{false} 
\def\pub{true} 
\input{template/packages/packages_general.tex}

\input{template/macros/macros_math.tex}

\input{template/macros/macros_general.tex}

\makeatletter
\def\bstctlcite{\@ifnextchar[{\@bstctlcite}{\@bstctlcite[@auxout]}}
\def\@bstctlcite[#1]#2{\@bsphack
  \@for\@citeb:=#2\do{%
    \edef\@citeb{\expandafter\@firstofone\@citeb}%
    \if@filesw\immediate\write\csname #1\endcsname
      {\string\citation{\@citeb}}\fi}%
  \@esphack}
\makeatother

\newcommand*{\Fyi}{{F_{y,i}}}
\newcommand*{\Fyf}{{F_{y,f}}}
\newcommand*{\Fyr}{{F_{y,r}}}
\newcommand*{\Fzi}{{F_{z,i}}}

\newcommand*{\delf}{{\delta_f}}
\newcommand*{\delr}{{\delta_r}}
\newcommand*{\lf}{{l_f}}
\newcommand*{\lr}{{l_r}}
\newcommand*{\alpi}{{\alpha_i}}
\newcommand*{\alpf}{{\alpha_f}}
\newcommand*{\alpr}{{\alpha_r}}
\newcommand*{\Ci}{{C_i}}
\newcommand*{\Cf}{{C_f}}
\newcommand*{\Cr}{{C_r}}
\newcommand*{\vx}{{v_x}}

\begin{document}
\bstctlcite{IEEEexample:BSTcontrol}

\title{
  Constrained Optimization-Based Yaw-Rate Reference Map Generation
  \\
  for Active Rear Steering Control of Four-Wheel Steering Vehicles
}

\author{
  Myeongseok Ryu${}^{1}$,
  Donghyun Hwang${}^{2}$,
  Youngsik Yoon${}^{2}$,
  and Kyunghwan Choi${}^{1*}$ }
\affils{ 
  ${}^{1}$Cho Chun Shik Graduate School of Mobility, Korea Advanced Institute of Science and Technology (KAIST), \\
Daejeon, 34051, Korea (\{dding\_98, kh.choi\}@kaist.ac.kr) {\small${}^{*}$ Corresponding author} \\
  ${}^{2}$Vehicle Motion Control Development Team, Hyundai Motor Company (HMC), \\
Hwaseong, 18280, Korea (\{hdh0914, ys\_yoon\}@hyundai.com) 
}



\abstract{
  The effectiveness of active rear steering (ARS) control for four-wheel steering (4WS) vehicles is widely recognized in the automotive industry.
  At low speeds, ARS control can enhance maneuverability by steering the rear wheels in the opposite direction to the front wheels, reducing the turning radius.
  In contrast, at high speeds, ARS control can improve stability by steering the rear wheels in the same direction as the front wheels, preventing oversteer behavior.
  However, the performance of ARS control is often limited by the reference model used to generate the desired yaw rate, which is typically derived from a steady state of front-wheel steering (FWS) vehicle model.
  In this paper, we conduct a numerical analysis to construct an optimal yaw rate reference map for ARS control by formulating a constrained optimization problem.
  In the optimization problem, constraints are imposed to ensure that the vehicle operates within safe limits at steady state.
  Numerical simulations demonstrate the effectiveness of the proposed method in providing an optimal yaw rate reference map for ARS control.
}


\maketitle


\section{Introduction}

The four-wheel steering (4WS) vehicles improve the conventional front-wheel steering (FWS) vehicles by adding the capability to steer the rear wheels, thereby enhancing vehicle maneuverability and stability.
In terms of control, the active rear steering (ARS) control had been widely studied over the past decades, particularly in the 1980s and 1990s.
However, the commercialization was limited due to the cost and complexity of the system \cite{Park:2025aa}.
Recently, owing to advance of electronic control technology, the ARS control has gained renewed attention, and several modern control strategies have been proposed in the literature \cite{Cho:1995aa,Bredthauer:2018aa,Eguchi:1989aa,Sano:1986aa,Guan:2024aa,Ro:1996aa,Akar:2006aa,Khanke:2019aa,Lv:2004aa,Hu:2023aa,Ryu:2026aa}.

Across the various control strategies, the core concept of ARS control is to generate larger yaw rate at low speeds by steering the rear wheels in the opposite direction to the front wheels, and to reduce sideslip angle at high speeds by steering the rear wheels in the same direction \cite{Cho:1995aa}.
These control behaviors result in improved maneuverability and enhanced stability at low and high speeds, respectively, which are the key performance metrics for vehicle lateral control.

Related works are reviewed including ARS and active four-wheel steering (A4WS) control methods.
Motivated by this concept, conventionally, many studies have adopted velocity-dependent ratio of rear steering angle to front steering angle, which is typically called proportional control \cite{Cho:1995aa,Bredthauer:2018aa}, \ie negative ratio at low speeds and positive ratio at high speeds.
In \cite{Eguchi:1989aa}, phase delay and phase reversal control methods were proposed, which mitigate the phase difference between yaw rate and lateral acceleration, and improve the transient response of the vehicle, respectively.
In addition, invoking the fact that front steering angle is typically limited to small values at high speeds, steering-angle-dependent ratio was proposed in \cite{Sano:1986aa}.
Regardless of vehicle velocity, the rear steering angle is steered in the same (opposite) direction to the front steering angle when the front steering angle is small (large).

Over the feedforward control, feedback control strategies also have been proposed.
Representatively, yaw rate tracking control methods have been developed, which control the rear steering angle to track a reference yaw rate, using various control techniques, such as proportional-integral-derivative (PID) control \cite{Guan:2024aa}, sliding mode control (SMC) \cite{Ro:1996aa,Akar:2006aa,Khanke:2019aa}, $H_\infty$ control \cite{Lv:2004aa}, and neuro-adaptive control (NAC) \cite{Ryu:2026aa}.
Particularly, some studies proposed zero-dynamics-based control methods, which assume that sideslip angle is zero and only consider yaw rate tracking control, since the sideslip angle is difficult to measure in real-time.
However, as reported in \cite{Hu:2023aa,Guan:2024aa}, the sideslip angle can be non-minimum phase, if the yaw rate is used as the only output of the system, when tire characteristics are saturated.
To address this issue, \cite{Guan:2024aa} proposed a control method that considers both yaw rate and sideslip angle in a weighted sum, and stable range of the weighting factor is proposed to ensure the stability of the system.

However, although the effectiveness of ARS control has been widely recognized and demonstrated, the performance of ARS control is often limited by the reference model used to generate the desired yaw rate, which is typically derived from FWS vehicle model at steady state \cite{Khanke:2019aa,Hu:2023aa}.
Hence, the fact that ARS control can improve yaw rate (maneuverability) while remaining the sideslip angle within safe limits (stability) is not fully exploited, which motivates the need for an optimal reference map for ARS control.
Thus, in this paper, we conduct a numerical analysis to construct an optimal yaw rate reference map for ARS control by formulating a constrained optimization problem.

This paper is organized as follows.
In Section~\ref{sec:vehicle:model}, the vehicle model used in the numerical analysis is introduced.
In Section~\ref{sec:numerical:analysis}, the constrained optimization problem for constructing the optimal reference map for ARS control is formulated, and the optimization results are analyzed.
In Section~\ref{sec:validation}, the optimization results were employed as the reference map for ARS control, and numerical simulations were conducted to validate the effectiveness of the proposed approach in improving the performance of ARS control for 4WS vehicles.
Finally, in Section~\ref{sec:conclusion}, the paper is concluded with a summary of the findings and future work.

\section{Vehicle Modelling} \label{sec:vehicle:model}

In this section, we introduce the vehicle model used for the numerical analysis.
The parameters of the vehicle model are listed in Table~\ref{table:notations}.

\begin{table}[t]
  \renewcommand{\arraystretch}{1.3}
  \caption{Vehicle Parameters.}
  \centering
  \small
  \begin{tabular}{c p{0.45\linewidth} c} 
  \hline
  \textbf{Symbol} & \textbf{Description} & \textbf{Value} \\
  \hline
  \hline 
  $m$ & Mass & \qty{2335.07}{\kilo\gram} \\
  \hline
  $h$ & Height of C.G. & \qty{0.42}{\meter} \\
  \hline
  $w$ & Track width & \qty{1.627}{\meter} \\
  \hline
  $\rho$ & Roll ratio & 0.6 \\
  \hline
  $I_z$  & Yaw moment of inertia & \qty{5376.432}{\kilo\gram\meter\squared} \\  
  \hline
  $\lf$ & Distance from C.G. to front axle & \qty{1.574}{\meter} \\
  \hline
  $\lr$ & Distance from C.G. to rear axle & \qty{1.566}{\meter} \\
  \hline
  $\Cf$ & Cornering stiffness of front tires & \qty{119.54}{\kilo\newton\per\radian} \\
  \hline
  $\Cr$ & Cornering stiffness of rear tires & \qty{119.83}{\kilo\newton\per\radian} \\
  \hline
  \end{tabular}
  \label{table:notations}
\end{table}

We adopt a 2-degree-of-freedom (2-DOF) bicycle model presented in \cite[Sec.~2.6]{Rajamani:2011aa}, as follows:
\begin{equation}\label{eq:2DOF:model}
  \ddt
  \begin{pmatrix}
    {\beta} \\
    {r}
  \end{pmatrix}
  =
  \begin{pmatrix}
    \tfrac{1}{m \vx} (\Fyf + \Fyr) - r \\
    \tfrac{1}{I_z} (\lf \Fyf - \lr \Fyr)
  \end{pmatrix}
  ,
\end{equation}
where $\beta$ and $r$ are the sideslip angle and yaw rate of the vehicle, respectively; $v_x$ is the longitudinal velocity of the vehicle; and $\Fyi$ is the lateral tire force at front ($i=f$) and rear ($i=r$) tires, respectively.

\begin{figure}[t]
  \centering
  \subfloat[Front lateral tire forces (245/46R19 at \qty{241}{\kilo\pascal}).]{
    \includegraphics[width=0.95\linewidth]{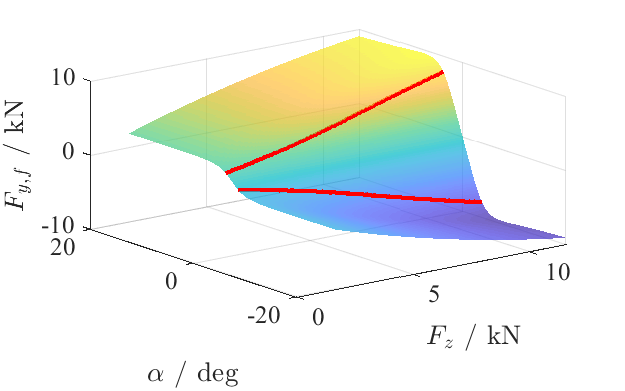}
    \label{fig:tire:forces:front}
  }
  \vfill
  \subfloat[Rear lateral tire forces (275/40R19 at \qty{241}{\kilo\pascal}).]{
    \includegraphics[width=0.95\linewidth]{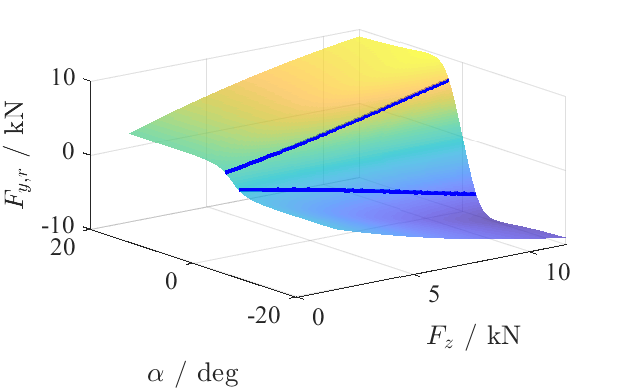}
    \label{fig:tire:forces:rear}
  }
  \caption{
    Lateral tire forces as functions of slip angles and normal forces.
    Red solid line \protect\colorLine{red}{solid} and blue solid line \protect\colorLine{blue}{solid} indicate bounds of the linear region of the tire forces for the front and rear tires, respectively.
  }
  \label{fig:tire:forces}
\end{figure}

The lateral tire forces are represented as nonlinear functions of the slip angles $\alpi$ and the normal forces $\Fzi$ for all $i\in\{f,r\}$.
We used the magic formula for the lateral tire forces in \cite[Sec.~13.5]{Rajamani:2011aa}, as follows:
\begin{equation} \label{eq:magic:formula}
  \begin{aligned}
    F_{y,i} 
    = &
    D_{y,i} \sin 
    \big[
      C_{y,i} \tan^{-1}
      \big\{
        B_{y,i} \widehat{\alpha}_{i}
        -
        E_{y,i} 
        \\ &
        \left( 
          B_{y,i} \widehat{\alpha}_{i}
          -
          \tan^{-1}
          (
            B_{y,i} \widehat{\alpha}_{i}
          ) 
        \right)
      \big\}
    \big]
    +
    S_{V,i}
    ,
  \end{aligned}
\end{equation}
where $\widehat{\alpha}_{i} := \alpi - S_{H,i}$.
The coefficients $B_{y,i}$, $C_{y,i}$, $D_{y,i}$, and $E_{y,i}$ are the stiffness, shape, peak, and curvature factors, and $S_{H,i}$ and $S_{V,i}$ are the horizontal and vertical shifts. 
Especially, the peak factor $D_{y,i}$ is a function of the normal force $\Fzi$ and road friction coefficient $\mu$ is defined as $0.9$ assuming dry asphalt road condition.
These coefficients are obtained from 245/46R19 and 275/40R19 tire models for the front and rear tires, respectively, under air pressure of $\qty{241}{\kilo\pascal}$.
The lateral tire forces are illustrated in Fig.~\ref{fig:tire:forces}, with respect to slip angles $\alpha$ and normal forces $F_z$. 
In addition, using small angle approximation, the slip angles are defined as follows:
\begin{equation}
  \alpf = \beta + \tfrac{\lf}{\vx} r - \delf 
  ,
  \quad
  \alpr = \beta - \tfrac{\lr}{\vx} r - \delr
  .
\end{equation}

\section{Numerical Analysis} \label{sec:numerical:analysis}

\subsection{Optimization Problem Formulation}

In this section, a constrained optimization problem for constructing the optimal yaw rate reference map for ARS control is formulated.
The vehicle parameters and dynamics used in the optimization problem are introduced in Section~\ref{sec:vehicle:model}.
The optimization problem is formulated as follows:
\begin{subequations} \label{eq:opt:prob} 
  \begin{align} 
    \min_{\delr} \quad 
    & 
    J(\delr)
    = 
    - r^2 + \lambda \beta^2
    \label{eq:opt:prob:obj} 
    \\ 
      \text{subject to} \quad 
    &
      \ddtt \beta = \ddtt r = 0,
    \label{eq:opt:prob:cstr:ss}
    \\ 
    & 
      \vert \beta \vert \le \overline{\beta},
    \label{eq:opt:prob:cstr:beta} 
    \\
    & 
      \vert a_{y} \vert \le \overline{a}_{y},
    \label{eq:opt:prob:cstr:acc} 
    \\
    &
      \vert \alpi \vert \le \overline{\alpha}_{i},
      \ \forall i \in \{f, r\}
    \label{eq:opt:prob:cstr:alpha}
    \\
    &
      \vert \delr \vert \le \overline{\delta}_r
    \label{eq:opt:prob:cstr:delr}
    ,
  \end{align}%
\end{subequations}
where $J:=J(\delr)$ is the objective function, $\lambda\in{\R}_{>0}$ denotes the weighting factor, and $\overline{\beta},\overline{a}_{y},\overline{\alpha}_{i},\forall i\in\{f,r\},\overline{\delta}_r\in{\R}_{>0}$ are the magnitude bounds for the sideslip angle, lateral acceleration, slip angles of front and rear tires, and rear steering angle, respectively.

The objective function \eqref{eq:opt:prob:obj} is designed to maximize the yaw rate and minimize the sideslip angle, which is a common performance metric for vehicle maneuverability and stability, respectively.
By tuning the weighting factor $\lambda$, the priority between maneuverability and stability can be adjusted.
In addition, the analysis is conducted under the steady-state condition, which is represented by the first constraint \eqref{eq:opt:prob:cstr:ss}.
The remaining constraints \eqref{eq:opt:prob:cstr:beta}--\eqref{eq:opt:prob:cstr:delr} are constraints that ensure the vehicle operates within safe limits at the steady-state condition, with $\overline{\beta}=\qty{3}{\deg}$, $\overline{a}_{y}=\qty{0.8}{\g}$, and $\overline{\delr}=\qty{3.5}{\deg}$.

Although the magic formula \eqref{eq:magic:formula} provides a more accurate description of nonlinear lateral tire forces, directly embedding it into the optimization problem \eqref{eq:opt:prob} may lead to a highly nonlinear and potentially ill-conditioned optimization problem.
Therefore, to ensure a well-posed and numerically tractable formulation, the vehicle dynamics are described using a linear tire model within the locally valid tire region, \ie $\Fyi = \Ci \alpi$, where $\Ci$ is the cornering stiffness obtained from the magic formula at $\ddtfrac{F_{y,i}}{\alpi}\big|_{\alpi=0}$ for all $i\in\{f,r\}$ as listed in Table~\ref{table:notations}.
The nonlinear tire characteristics are instead incorporated through linear slip angle constraints, whose admissible bounds are determined from the magic formula in \eqref{eq:opt:prob:cstr:alpha}, which is illustrated in Fig.~\ref{fig:tire:forces} as red solid line and blue solid line for the front and rear tires, respectively, \ie 10\% deviation from the linear region is allowed.

Moreover, considering lateral acceleration at steady state ($a_y = \vx r$), the normal forces are calculated using static weight distribution \cite[Sec.~18.3]{Milliken:1996aa}, as follows:
\begin{equation}
  \begin{pmatrix}
    F_{z,fl}
    \\
    F_{z,fr}
    \\
    F_{z,rl}
    \\
    F_{z,rr}
  \end{pmatrix}
  =
  \begin{pmatrix}
    \tfrac{mg\lr}{2(\lf+\lr)} 
    \\
    \tfrac{mg\lr}{2(\lf+\lr)} 
    \\
    \tfrac{mg\lf}{2(\lf+\lr)} 
    \\
    \tfrac{mg\lf}{2(\lf+\lr)} 
  \end{pmatrix}
  +
  \begin{pmatrix}
    -\rho\tfrac{m a_y h}{2w}
    \\
    +\rho\tfrac{m a_y h}{2w}
    \\
    -(1-\rho)\tfrac{m a_y h}{2w}
    \\
    +(1-\rho)\tfrac{m a_y h}{2w}
  \end{pmatrix}
  ,
\end{equation}
where subscripts $fl$, $fr$, $rl$, and $rr$ denote the front-left, front-right, rear-left, and rear-right tires, respectively; and $g=9.81$ is the gravitational acceleration.

\subsection{Analysis Results}


\begin{table}[t]
  \renewcommand{\arraystretch}{1.3}
  \caption{Configurations of Optimization Problem.}
  \centering
  \small
  \begin{tabular}{c l} 
  \hline
    & \textbf{Optimization Configuration} \\
  \hline
  \hline 
      FWS
    & 
    Front wheel steering only ($\delr = 0$, $\lambda = 100$)
  \\
  \hline 
    \multirow{2}{*}{
      4WS-1
    }
    & 
    Four-wheel steering with
    \\
    &
    low penalty on sideslip angle ($\lambda = 100$)
  \\
  \hline 
  \multirow{2}{*}{
    4WS-2 
  }
    & 
    Four-wheel steering with
    \\
    &
    high penalty on sideslip angle ($\lambda = 3000$) 
  \\
  \hline
  \end{tabular}
  \label{table:opt:cmp:cases}
\end{table}

\begin{figure*}[t]
    \centering
    \subfloat[FWS configuration ($\delta_r=0, \lambda=100$).]{
      \includegraphics[width=0.99\linewidth]{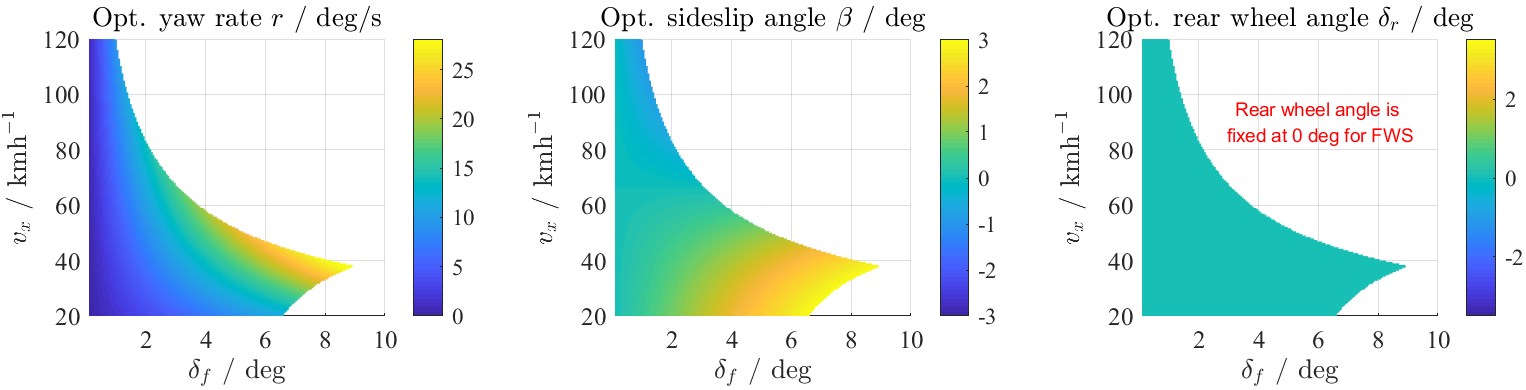}
      \label{fig:opt:results:FWS}
    }
    \vfill
    \subfloat[4WS-1 configuration with low beta penalty ($\lambda=100$).]{
      \includegraphics[width=0.99\linewidth]{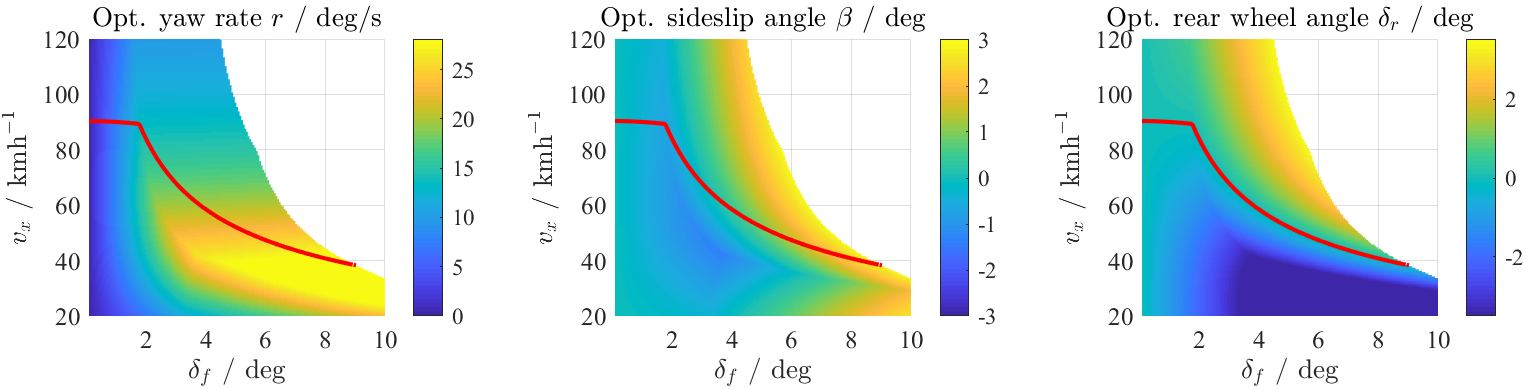}
      \label{fig:opt:results:ARS-1}
    }
    \vfill
    \subfloat[4WS-2 configuration with high beta penalty ($\lambda=3000$).]{
      \includegraphics[width=0.99\linewidth]{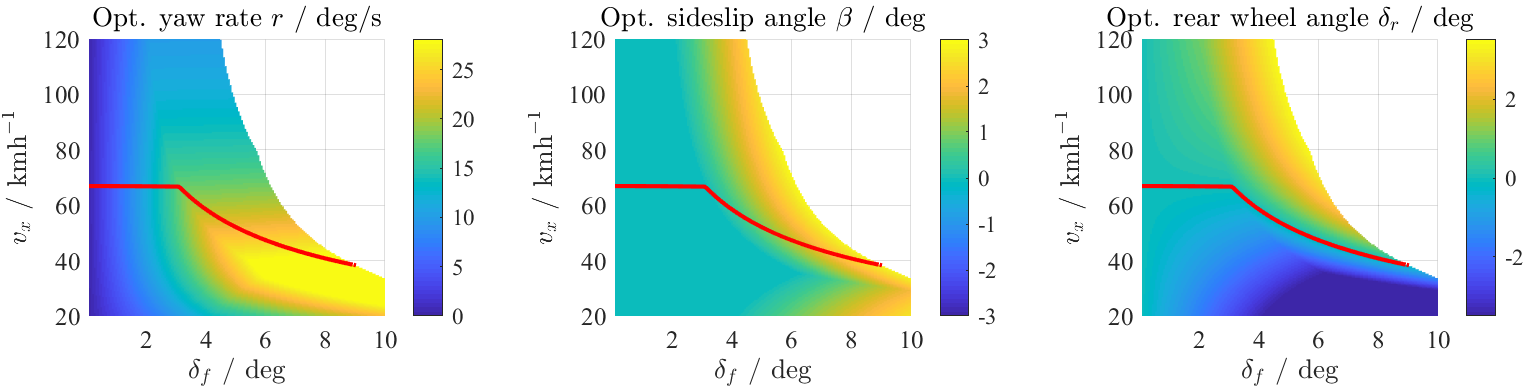}
      \label{fig:opt:results:ARS-2}
    }
    \caption{
      Optimization results of \eqref{eq:opt:prob} for various longitudinal velocities $\vx$ and front steering angles $\delf$.
      Red solid line \protect\colorLine{red}{solid} indicates phase conversion from opposite and same directional steering of rear wheels to front wheels.
    }
\label{fig:opt:results}
\end{figure*}

\begin{figure}[t]
  \centering
  \subfloat[Optimal yaw rates $r$.]{
    \includegraphics[width=0.98\linewidth]{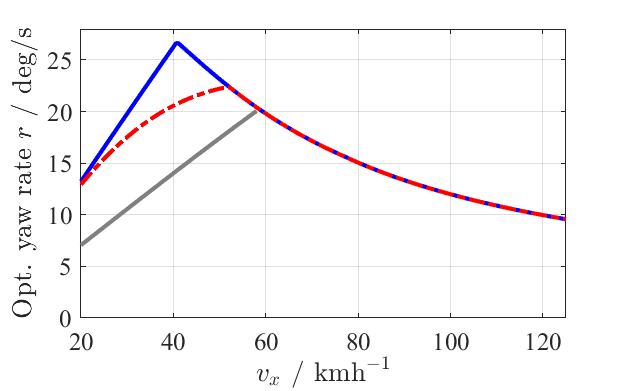}
    \label{fig:opt:results:specific:uf:yaw:rate}
  }
  \vfill
  \subfloat[Optimal sideslip angles $\beta$.]{
    \includegraphics[width=0.98\linewidth]{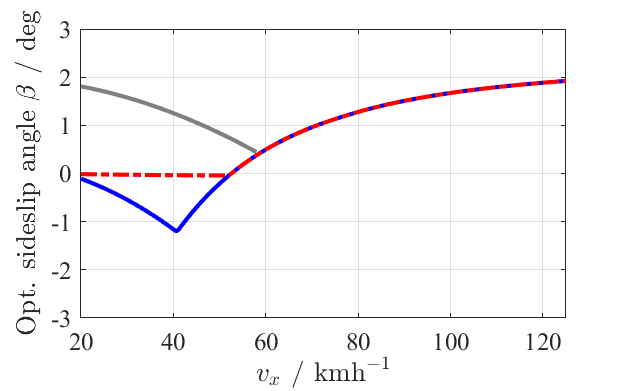}
    \label{fig:opt:results:specific:uf:side:slip}
  }
  \caption{
    Optimal yaw rates and sideslip angles at $\delf = \qty{4.0}{\deg}$ of FWS  \protect\colorLine{gray}{solid} and 4WS-1 \protect\colorLine{blue}{solid}, and 4WS-2 \protect\colorLine{red}{dashed}.
  }
  \label{fig:opt:results:specific:uf}
\end{figure}

The optimization problem \eqref{eq:opt:prob} is solved for various longitudinal velocities $\vx \in \{20,\cdots,120\}$ \qty{}{\kilo\meter\per\hour}, and front wheel angles $\delf \in \{0.1,\cdots,10\}$ \qty{}{\deg}.
Three configurations are compared, as listed in Table~\ref{table:opt:cmp:cases}.
The optimization results are illustrated in Figs.~\ref{fig:opt:results}.
Specifically, the optimization results at $\delf = \qty{4.0}{\deg}$, are highlighted in Fig.~\ref{fig:opt:results:specific:uf} with quantitative comparison in Table~\ref{table:opt:cmp:specific:uf}.

In Fig.~\ref{fig:opt:results:FWS}, the optimization results for the FWS configuration, show that the operating domain, $\vx$ and $\delf$, is limited since there is no feasible solution which satisfies the imposed constraints.
Moreover, since the rear wheels are not steered, the sideslip angles are passively determined by the front wheel steering angle.

In contrast, in Figs.~\ref{fig:opt:results:ARS-1} and \ref{fig:opt:results:ARS-2}, the optimization results for the 4WS configurations show that the operating domain is significantly expanded, and the optimal yaw rates are improved compared to the FWS configuration, as well.
This demonstrates the effectiveness of ARS control in enhancing the maneuverability of 4WS vehicles, ensuring the vehicle operates within safe limits at steady state.
Specifically, at \qty{40}{}--\qty{45}{\kilo\meter\per\hour}, the optimal yaw rates are significantly improved when the front wheel angle is $\qty{4.0}{\deg}$, as shown in Fig.~\ref{fig:opt:results:specific:uf:yaw:rate}.

The effect of the weighting factor $\lambda$ in the objective function \eqref{eq:opt:prob:obj} is also observed in Figs.~\ref{fig:opt:results:ARS-1} and \ref{fig:opt:results:ARS-2}.
By increasing $\lambda$, the optimal yaw rates are reduced, but the optimal sideslip angles are around zero in wide operating domain, by steering the rear wheels in the same direction to the front wheels.
It is notable that red solid line in Figs.~\ref{fig:opt:results:ARS-1} and \ref{fig:opt:results:ARS-2} indicates the phase conversion from opposite direction to same direction of rear wheels to the front wheels, which implies that higher $\lambda$ leads to earlier phase conversion with respect to longitudinal velocity and front wheel angle.

Additionally, the ideal rear steering strategies at steady state for the 4WS configurations can be observed.
As shown in Figs.~\ref{fig:opt:results:ARS-1} and \ref{fig:opt:results:ARS-2}, the optimal rear steering angles $\delr$ are negative at low speeds and positive at high speeds, indicating that steering the rear wheels in the opposite and same direction to the front wheels is optimal for enhancing maneuverability and stability, respectively.

\hfill

We detail the comparison of the optimization results, especially at $\delf = \qty{4.0}{\deg}$.
As shown in Fig.~\ref{fig:opt:results:specific:uf}, the yaw rates of the 4WS configurations are significantly improved compared to the FWS configuration, while the sideslip angles are maintained within the safety limits.
Quantitatively, as shown in Table~\ref{table:opt:cmp:specific:uf}, the maximum yaw rate is improved by 74.7\% and 38.1\% for 4WS-1 and 4WS-2 configurations, respectively, compared to the FWS configuration, while the minimum absolute value of the sideslip angle is reduced by 1.9\% and 97.3\%, respectively, at $\vx = \qty{43.9}{\kilo\meter\per\hour}$.
Moreover, by increasing $\lambda$ from 4WS-1 to 4WS-2, the maximum yaw rates are reduced, but the sideslip angles are significantly reduced; specifically, the sideslip angles are almost zero when $20 \le \vx \le 55$ \qty{}{\kilo\meter\per\hour}.

\begin{table}[t]
  \renewcommand{\arraystretch}{1.3}
  \caption{Optimization results at $\delf = \qty{4.0}{\deg}$, and $\vx = \qty{43.9}{\kilo\meter\per\hour}$.}
  \centering
  \small
  \begin{tabular}{c c c} 
  \hline
    & \textbf{Max. yaw rate $r$} & \textbf{Min. sideslip angle $\vert\beta\vert$} \\
  \hline
  \hline 
      FWS
    &
      \qty{15.35}{\deg/\second}  (-) 
    &
      \qty{1.101}{\deg} (-)
  \\
  \hline 
    4WS-1
    &
      \qty{26.82}{\deg/\second} (+74.7\%) 
    &
      \qty{1.08}{\deg} (-1.9\%)
  \\
  \hline
    4WS-2 
    &
      \qty{21.35}{\deg/\second} (+38.1\%) 
    &
      \qty{0.03}{\deg} (-97.3\%)
    \\
    \hline 
  \end{tabular}
  \label{table:opt:cmp:specific:uf}
\end{table}

\section{Numerical Validation} \label{sec:validation}

\subsection{Validation Setup}

In this section, a numerical validation was conducted to evaluate the effectiveness of the proposed method by employing the optimization results as a yaw rate reference map for ARS control of 4WS vehicles.

The vehicle parameters used in the simulations are the same as those used in Section~\ref{sec:numerical:analysis}.
Test scenario was designed as step steering inputs at $\vx = \qty{42.5}{\kilo\meter\per\hour}$.
The driver's command was step steering input from \qty{0}{\deg} to $\qty{90}{\deg}$ at $t = \qty{10}{\second}$, under $\qty{300}{\deg/\second}$ rate limit of the steering wheel.
The steering wheel angle of $\qty{90}{\deg}$ corresponds to the front wheel angle $\delf$ of $\qty{6.75}{\deg}$.
The road condition was assumed to be dry asphalt, and the road friction coefficient was set to $\mu = 0.9$.

The yaw rate tracking controller for ARS control was designed as a simple proportional-integral (PI) controller, as follows:
\begin{equation}
  \delr(t) = K_p (r_{\mathrm{ref}}(t) - r(t)) + K_i \int_0^t (r_{\mathrm{ref}}(\tau) - r(\tau)) \der\tau
  ,
\end{equation}
where $r_{\mathrm{ref}}(t)$ is the reference yaw rate obtained from the optimization results, and $K_p=0.01$ and $K_i=1$ are the proportional and integral gains, respectively.
The reference yaw rate was obtained by linearly interpolating the optimal yaw rate of the optimization results for the given longitudinal velocity and front steering angle.

The simulation was implemented in IPG CarMaker 15~\cite{IPGCarMaker:2026aa}.
The sampling time was set to $\qty{1}{\milli\second}$ for simulation, and $\qty{10}{\milli\second}$ for the controllers.

For comparative study, the reference map obtained from 4WS-2 configuration was used for (C$_1$), and the reference map obtained from FWS configuration was used for (C$_2$).

\subsection{Validation Results}

\begin{figure}[t]
  \subfloat[Trajectory of yaw rate $r$.]{
    \includegraphics[width=0.98\linewidth]{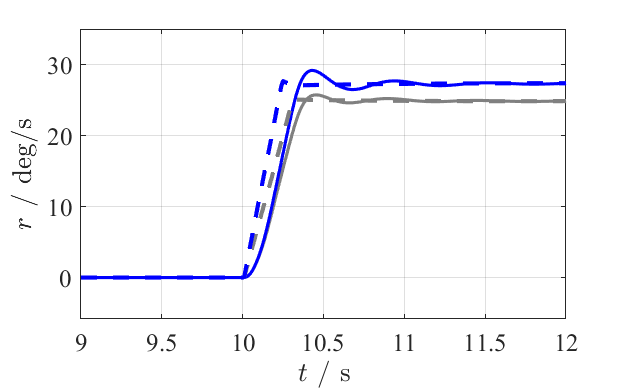}
    \label{fig:sim:results:yaw:rate}
  }
  \vfill
  \subfloat[Trajectory of sideslip angle $\beta$.]{
    \includegraphics[width=0.98\linewidth]{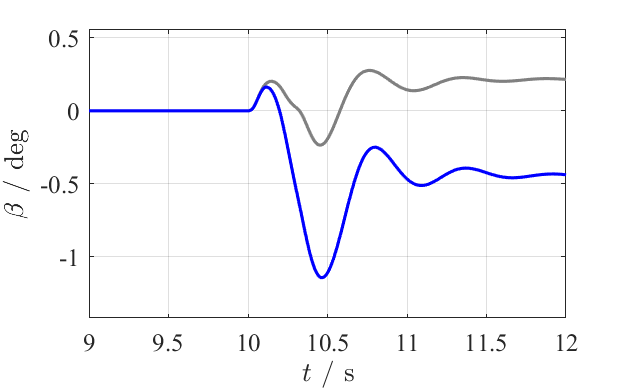}
    \label{fig:sim:results:beta}
  }
  \vfill
  \subfloat[Trajectory of rear steering angle $\delr$.]{
    \includegraphics[width=0.98\linewidth]{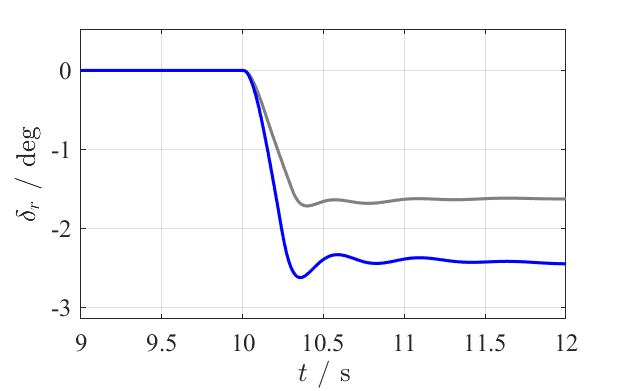}
    \label{fig:sim:results:ur}
  }
  \caption{
    Simulation results of (C$_1$) \protect\colorLine{blue}{solid} and (C$_2$) \protect\colorLine{gray}{solid} at $\vx = \qty{42.5}{\kilo\meter\per\hour}$ and $\qty{90}{\deg}$ step steering scenarios.
    Yaw rate references are illustrated as \protect\colorLine{blue}{dashed} and \protect\colorLine{gray}{dashed} for (C$_1$) and (C$_2$), respectively.
  }
  \label{fig:sim:results}
\end{figure}

The simulation results are illustrated in Fig.~\ref{fig:sim:results} and quantitatively compared in Table~\ref{table:sim:results}.
As shown in Fig.~\ref{fig:sim:results}, (C$_1$) utilizes better capability of ARS control to generate larger yaw rate while maintaining the sideslip angle within the safety limits, \ie $\vert\beta\vert \le \qty{3}{\deg}$.
In contrast, the yaw rate of (C$_2$) is lower than that of (C$_1$) around 9.2\% at the steady state, as shown in Table~\ref{table:sim:results}.
This is because the target yaw rate of (C$_2$) is generated assuming the ARS control is unavailable (FWS configuration), resulting in a lower target yaw rate compared to (C$_1$).

It should be noted that the sideslip angle of (C$_1$) is larger than that of (C$_2$).
This demonstrates the effect of ARS control in opposite-directional rear steering.
Increasing the magnitude of opposite-directional rear steering, the sideslip angle is biased toward negative direction.
Considering this effect, the weighting factor $\lambda$ in the optimization problem \eqref{eq:opt:prob} should be carefully tuned.
Smaller $\lambda$ in the optimization problem may lead to larger optimal reference yaw rate, but it may also lead to larger sideslip angle, which can affect the stability of the vehicle.
On the other hand, larger $\lambda$ may lead to more safe reference yaw rate, but it may reduce the maneuverability of the vehicle by reducing the reference yaw rate.


\begin{table}[t]
  \renewcommand{\arraystretch}{1.3}
  \caption{Quantitative Comparison of Simulation Results.}
  \centering
  \small
  \begin{tabular}{p{0.2\linewidth} c c} 
  \hline
    & \textbf{Max. Yaw Rate $r$} & \textbf{Min. Sideslip Angle $\vert\beta\vert$} \\
  \hline
  \hline 
      (C$_1$) \protect\colorLine{blue}{solid}
    &
      \qty{27.39}{\deg/\second}
    &
      \qty{0.44}{\deg} 
  \\
  \hline 
      (C$_2$) \protect\colorLine{gray}{solid}
    &
      \qty{24.86}{\deg/\second}
    &
      \qty{0.22}{\deg}
    \\
    \hline 
  \end{tabular}
  \label{table:sim:results}
\end{table}

\section{Conclusion} \label{sec:conclusion}

In this paper, we analyzed the effectiveness of ARS control for 4WS vehicles by formulating a constrained optimization problem to construct the optimal yaw rate reference map.
The optimization problem was designed to maximize the yaw rate of the vehicle while minimizing the sideslip angle, subject to constraints that ensure the vehicle operates within safe domains.
The numerical analysis with the optimization results demonstrated that ARS control can significantly expand the feasible operating regions and improve the feasible maximum yaw rates.

Moreover, the optimization results were employed as a yaw rate reference map for ARS control in numerical simulations.
The numerical simulations demonstrated the effectiveness of the proposed method in improving yaw rate while maintaining the sideslip angle within safety limits.

As future work, the reference map will be employed in advanced ARS control strategies, such as model predictive control (MPC) and real-time implementation on Electronic Control Units (ECUs) for real-world testing and validation.

\section*{ACKNOWLEDGEMENT}

This work was supported in part by Hyundai Motor Company (HMC).

\bibliographystyle{IEEEtran}
\bibliography{refs}

\end{document}

%% file: template/packages/packages_general.tex
\usepackage{balance}

\usepackage{siunitx}    
\usepackage{kotex}      
\usepackage{lipsum}     
\usepackage{array}

\usepackage{cite}   

\usepackage{amsmath,amsfonts}
\usepackage{amssymb}
\usepackage{soul}

\def\publish{true}

\usepackage{hyperref}
\ifx \pub\publish 
    \hypersetup{
        pdftoolbar=false,        	
        pdfmenubar=false,        	
        pdffitwindow=false,     	
        pdfstartview={FitH},    	
        pdftitle={},    	
        pdfauthor={},     	
        colorlinks=true,       	
        linkcolor=black,          	
        citecolor=black,       	
        filecolor=black,      	
        urlcolor=black,		
        backref=page,            
    }
\else 
    \hypersetup{
        pdftoolbar=false,        	
        pdfmenubar=false,        	
        pdffitwindow=false,     	
        pdfstartview={FitH},    	
        pdftitle={},    	
        pdfauthor={},     	
        colorlinks=true,       	
        linkcolor=blue,          	
        citecolor=blue,       	
        filecolor=blue,      	
        urlcolor=blue,		
        backref=page,            
    }
\fi

\usepackage{algorithm}
\usepackage{algorithmic}

\usepackage{epsfig}
\usepackage[dvipsnames]{xcolor}

\usepackage{svg}
\usepackage{subfig}

\usepackage{textcomp}
\usepackage{stfloats}
\usepackage{url}
\usepackage{verbatim}
\usepackage{graphicx}
\usepackage{multirow}
\usepackage{multicol}

\usepackage{amsthm}
\usepackage{lipsum}
\usepackage{tikz}

%% file: template/macros/macros_math.tex
\newcommand\R{\mathbb{R}}

\newcommand\der{\mathrm d}
\newcommand*{\ddt}{
    \frac{\der}{\der t}
}
\newcommand*{\ddtt}{
    \tfrac{\der}{\der t}
}

\newcommand*{\ddtfrac}[2]{
    \tfrac{\der {#1}}{\der {#2}}
}


%% file: template/macros/macros_general.tex
\def\comment{true}
\usepackage{soul}
\definecolor{airforceblue}{rgb}{0.36, 0.54, 0.66}
\definecolor{awesome}{rgb}{1.0, 0.13, 0.32}
\definecolor{blush}{rgb}{0.87, 0.36, 0.51}
\definecolor{my_cyan}{rgb}{0.0, 1.0, 1.0}
\definecolor{darklavender}{rgb}{0.45, 0.31, 0.59}
\definecolor{cinnabar}{rgb}{0.89, 0.26, 0.2}

\newcommand{\KHCH}[1]{
    \ifx \cmt\comment{\color{RoyalPurple} [KH: #1]}\else{}\fi
} 
\newcommand{\MSRY}[1]{
    \ifx \cmt\comment{\color{Peach} [MS: #1]}\else{}\fi
} 
\newcommand{\SHJH}[1]{
    \ifx \cmt\comment{\color{green} [SH: #1]}\else{}\fi
} 
\newcommand{\DHHO}[1]{
    \ifx \cmt\comment{\color{magenta} [DH: #1]}\else{}\fi
} 
\newcommand{\JYKI}[1]{
    \ifx \cmt\comment{\color{blush} [JY: #1]}\else{}\fi
} 
\newcommand{\NSER}[1]{
    \ifx \cmt\comment{\color{cinnabar} [NS: #1]}\else{}\fi
} 
\newcommand{\SBHW}[1]{
    \ifx \cmt\comment{\color{darklavender} [SB: #1]}\else{}\fi
} 

\newcommand\ie{\textrm{i.e.,\ }}

\newcommand{\colorLine}[2]{[\tikz[baseline=(current bounding box.base)]{\draw[color=#1,
#2,line width=2pt] (0,3pt) -- (1.5em,3pt);}]}
